\documentclass{pasj00}
\begin{document}
\SetRunningHead{Mukai et al.}{X-ray Decline of CH Cyg}
\Received{2006/07/19}
\Accepted{2006/08/21}

\title{An Apparent Hard X-ray Decline of CH Cygni}

%%% begin:list of authors
%\author{Koji \textsc{Mukai}%
%  \thanks{Also Universities Space Research Association}}
%\affil{Code 662, NASA/GSFC, Greenbelt, MD 20771, USA}\email{mukai@milkyway.gsfc.nasa.gov}
%
%\author{Manabu \textsc{Ishida}}
%\affil{Department of Physics, Tokyo Metropolitan University, 1-1 Minami-Ohsawa, Hachioji, Tokyo 192-0397}\email{ishida@mpuax.phys.metro-u.ac.jp}
%
%\author{Caroline {\sc Kilbourne}}
%\affil{Code 662, NASA/GSFC, Greenbelt, MD 20771, USA}\email{cak@lheapop.gsfc.nasa.gov}
%
%\author{Hideyuki {\sc Mori}}
%\affil{Department of Physics, Graduate School of Science, Kyoto University, Sakyo-ku, Kyoto 606-8502}\email{mori@cr.scphys.kyoto-u.ac.jp}
%
%\author{Yukikatsu {\sc Terada}}
%\affil{Cosmic Radiation Laboratory, RIKEN, 2-1 Hirosawa, Wako, Saitama 351-0198 }\email{terada@riken.jp}
%
%\and
%
%\author{Others from {\sc the Suzaku Team}}
%\affil{D-Address of Institute}\email{ddddd@xxx.xxx.xx.xx}

%%% end:list of authors

%%% Please use the following style in case that sorting by 
%%% affilation is impossible. 
%
%%% begin:list of authors

\author{%
    Koji \textsc{Mukai}\altaffilmark{1,2}
    Manabu \textsc{Ishida}\altaffilmark{3}
    Caroline \textsc{Kilbourne}\altaffilmark{1}
    Hideyuki \textsc{Mori}\altaffilmark{4}}
\author{
    Yukikatsu \textsc{Terada}\altaffilmark{5}
    Kai-Wing \textsc{Chan}\altaffilmark{1,2} \and
    Yang \textsc{Soong}\altaffilmark{1,2}}

\altaffiltext{1}{Code 662, NASA/GSFC, Greenbelt, MD 20771, USA}
\email{mukai@milkyway.gsfc.nasa.gov}
\email{cak@lheapop.gsfc.nasa.gov}
\email{kwchan@milkyway.gsfc.nasa.gov}
\email{soong@milkyway.gsfc.nasa.gov}
\altaffiltext{2}{Also Universities Space Research Association}
\altaffiltext{3}{Department of Physics, Tokyo Metropolitan University, 1-1 Minami-Ohsawa, Hachioji, Tokyo 192-0397}
\email{ishida@mpuax.phys.metro-u.ac.jp}
\altaffiltext{4}{Department of Physics, Graduate School of Science, Kyoto University, Sakyo-ku, Kyoto 606-8502}
\email{mori@cr.scphys.kyoto-u.ac.jp}
\altaffiltext{5}{Cosmic Radiation Laboratory, RIKEN, 2-1 Hirosawa, Wako, Saitama 351-0198}
\email{terada@riken.jp}

%% `\KeyWords{}' always has to be placed before `\maketitle'.
 %Do NOT move this preamble from here!

\maketitle

\begin{abstract}
CH Cygni is a symbiotic star consisting of an M giant and an accreting
white dwarf, which is known to be a highly variable X-ray source with
a complex, two-component, spectra.  Here we report on two {\sl Suzaku\/}
observations of CH Cyg, taken in 2006 January and May, during which the
system was seen to be in a soft X-ray bright, hard X-ray faint state.
Based on the extraordinary strength of the 6.4 keV fluorescent Fe
K$\alpha$ line, we show that the hard X-rays observed with {\sl Suzaku\/}
are dominated by scattering.
\end{abstract}

\section{Introduction}

A symbiotic star is a binary composed of a red giant and a hot blue
component, often an accreting white dwarf.  Many symbiotic stars
have been detected as soft X-ray sources using {\sl ROSAT\/}
\citep{MWJ1997}.  The {\sl ROSAT\/} detections can usually be classified
into two types.  One is of the supersoft type, which can be understood
as photospheric emission from the hot white dwarf.  The second type
is optically thin plasma emission with temperatures of order 10$^7$K,
which \authorcite{MWJ1997} attributed to the colliding winds.
That is, a strong, fast ($\sim$1000 km\,s$^{-1}$) wind from the
vicinity of the accreting white dwarf, which, in many cases, are
known to be present through studies of UV and optical emission lines,
is colliding with the dense, low-velocity wind from the M giant,
shock-heating the gas to X-ray emitting temperatures.  More recently,
a subset of symbiotic stars have been shown to be a strong source of
hard ($>$10 keV) X-rays in the {\sl INTEGRAL\/} Galactic plane scan
and in the {\sl Swift\/} BAT survey (\cite{Kea2006} and references
therein), which requires the presence of yet another type of X-ray
emission in these symbiotic stars.

CH Cygni is an oft-observed, yet poorly-understood symbiotic star,
at a distance of 245$\pm$50 pc according to the {\sl Hipparcos\/}
parallax measurements \citep{Pea1997}.  It has produced radio
\citep{TSM1986} and optical \citep{S1987} jets, which appear to
be ejected in the plane of the sky as seen from the Earth,
implying an accretion disk and/or binary with an inclination
angle of nearly 90$^\circ$.  \citet{Hea1993} analyzed a large
body of infrared spectra and proposed CH Cyg to be a triple system,
consisting of a 756 day period inner binary in
an 14.5 yr orbit around a third component.  However, \citet{Mea1996}
cautioned against the triple system interpretation on several grounds.
In particular, the orbital solution of \authorcite{Hea1993} requires
a low inclination angle, contrary to the indications from the jet data.
In addition, such a low inclination then results in the unrealistically
low mass (0.2M$_\odot$) for the white dwarf.  To add to the confusion,
\citet{Sea1996} claimed that CH Cyg was a triple system as proposed by
\citet{Hea1993} but both the inner and the outer binaries were eclipsing,
thus requiring high inclination angles for both.  More recently, the
similarity between the 756 day period of CH Cyg with non-radial
pulsations in quite a few semi-regular variables has become apparent;
therefore, the 756 day period of CH Cyg is likely to be a non-radial
pulsation mode rather than an orbital period (\cite{Sea2006} and
references therein).

CH Cyg has been known to be a variable X-ray source for many years;
the early X-ray data from {\sl HEAO-1\/} to {\sl EXOSAT\/} are
summarized by \citet{LT1987}.  It was detected with {\sl ROSAT\/}
showing the colliding wind-type spectrum and an additional hard
component \citep{MWJ1997}.  However, it was the {\sl ASCA\/} observation
that revealed the dramatic two-component X-ray spectrum \citep{EIM1997}.
There is a relatively unabsorbed soft component, dominant below 2 keV,
that can be modeled as two-temperature (kT$\sim$0.2, 0.7 keV) optically
thin thermal plasma emission.  Above 2 keV, there is a separate, highly
absorbed hotter component (kT$\sim$7 keV).  \citet{EIM1997} interpreted
the latter as due to accretion onto a $>$0.44 M$_\odot$ white dwarf,
and the former as either from the M giant or from the jets.  \citet{W2001}
proposed an alternative interpretation of a single spectral component
seen through an ionized wind of the M giant.  A {\sl Chandra\/} HETG
observation was performed to decide between the two models, but CH Cyg
was found to be much fainter than during the {\sl ASCA\/} observation,
making the grating data of little use.  Instead, \citet{GS2004} analyzed
the 0th order data and discovered a weak, spatially extended emission
from the jet, in addition to the much brighter, spatially unresolved,
component.

Here we report on two {\sl Suzaku\/} observations of CH Cyg performed
in 2006.  We have also re-analyzed the {\sl ASCA\/} data and
the {\sl Chandra\/} zeroth order data, and present the results in
a uniform manner.

\section{Suzaku Observations}

We observed CH Cyg twice with the Japanese-US X-ray astronomy
satellite, {\sl Suzaku\/} \citep{SUZ2006}.  The exposure times
with the X-ray Imaging Spectrometer (XIS) instrument \citep{XIS2006}
was approximately 35 ksec on both occasions. We have analyzed the data
prepared using the version 0.7 pipeline \citep{SUZ2006}.  This results
in an energy scale accurate to $\pm$0.2\% at Fe K$\alpha$ and
$\pm$5 eV below 1 keV \citep{XIS2006}.  Although CH Cyg is definitely
detected with the PIN detector of the Hard X-ray Detector (HXD;
\cite{HXD2006}) instrument at approximately
7$\times 10^{-12}$ ergs\,cm$^{-2}$s$^{-1}$ in the 12--25 keV band,
this is sufficiently close to the current systematic uncertainties
in the background estimation \citep{HXDPER2006} that we have chosen
to concentrates on the XIS data.  We have extracted the XIS
data from a 2.5 arcmin radius extraction region centered on the source,
used the 2.5--5.0 arcmin annulus as the background, and generated
response files taking into account the vignetting of the X-Ray Telescopes
(XRT; \cite{XRT2006}) and the time-variable contamination on the
optical blocking filter of the XIS\footnote{The calibration of the
contamination layer is ongoing, and may result in slight errors
particularly for the May 2006 observation, in part because the calibration
is based on version 0.6 processing.  We do not believe this causes
a serious problem in the context of this paper.}.  We have combined
the spectra taken with the three units (XIS0, XIS2, and XIS3) with
frontside illuminated (FI) CCDs, and analyze the combined spectrum
simultaneously with that taken with XIS1 with the backside illuminated
(BI) CCD.   Our source extraction region contains roughly 93\% of the
source flux, and this is taken into account by the response file.
On the other hand, it does not account for the fact that the
remaining 7\% of the flux is in the background region\footnote{The fraction
of photons that fall outside the 5.0 arcmin outer radius of the
background region is negligibly small \citep{XRT2006}.}, and this has
been subtracted away; we have therefore multiplied the fluxes derived
from the spectral fits by 1.08 ($=93/(93-7)$).  The log of
{\sl Suzaku\/} observations can be found in Table\,\ref{tab:obslog},
together with a summary of the {\sl ASCA\/} (see \cite{EIM1997} for details)
and {\sl Chandra\/} (see \cite{GS2004} for details) data.

\begin{table*}
  \caption{Log of S{\sl Suzaku\/} Observations.}\label{tab:obslog}
  \begin{center}
    \begin{tabular}{llllll}
      Satellite & Sequence No.& Date of Obs. & Exposure & 0.4--2 keV flux & 3--10 keV flux\\
       & &  & (ksec) & (10$^{-12}$ ergs\,cm$^{-2}$s$^{-1}$) & (10$^{-12}$ ergs\,cm$^{-2}$s$^{-1}$) \\
      {\sl ASCA\/} & 42020000 & 1994 Oct 19 & 21 & 3.3 & 63.4 \\
      {\sl Chandra\/} & 1904 & 2001 Mar 27 & 47 & 0.4 & 1.4 \\
      {\sl Suzaku\/} & 400016020 & 2006 Jan 04/05 & 35 & 3.8 & 2.1 \\
      {\sl Suzaku\/} & 400016030 & 2006 May 28/29 & 35 & 2.7 & 2.0 \\
    \end{tabular}
  \end{center}
\end{table*}

\section{Results}

\begin{figure*}
  \begin{center}
    \FigureFile(123.69375mm,78.84mm){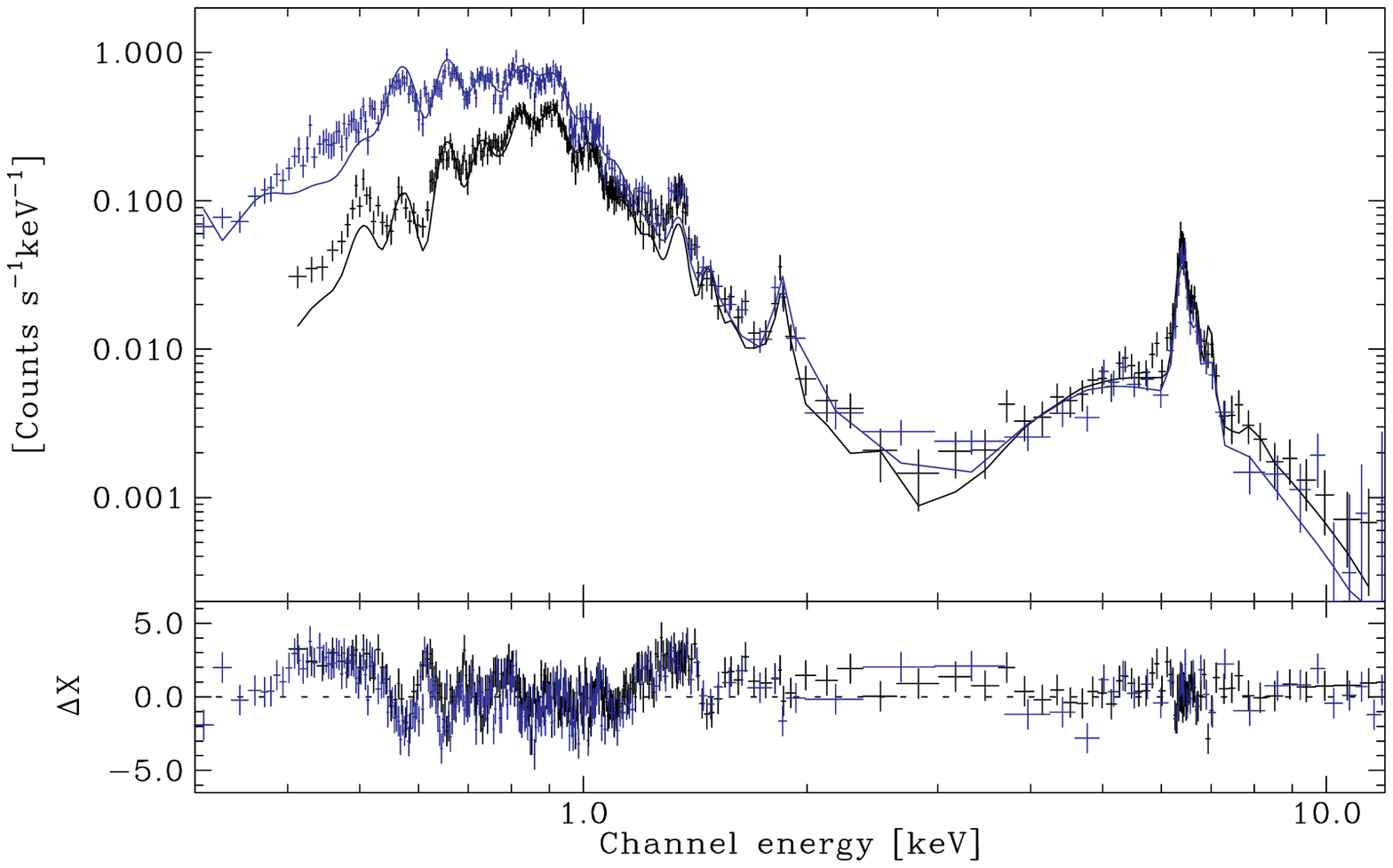}
    \FigureFile(123.69375mm,78.84mm){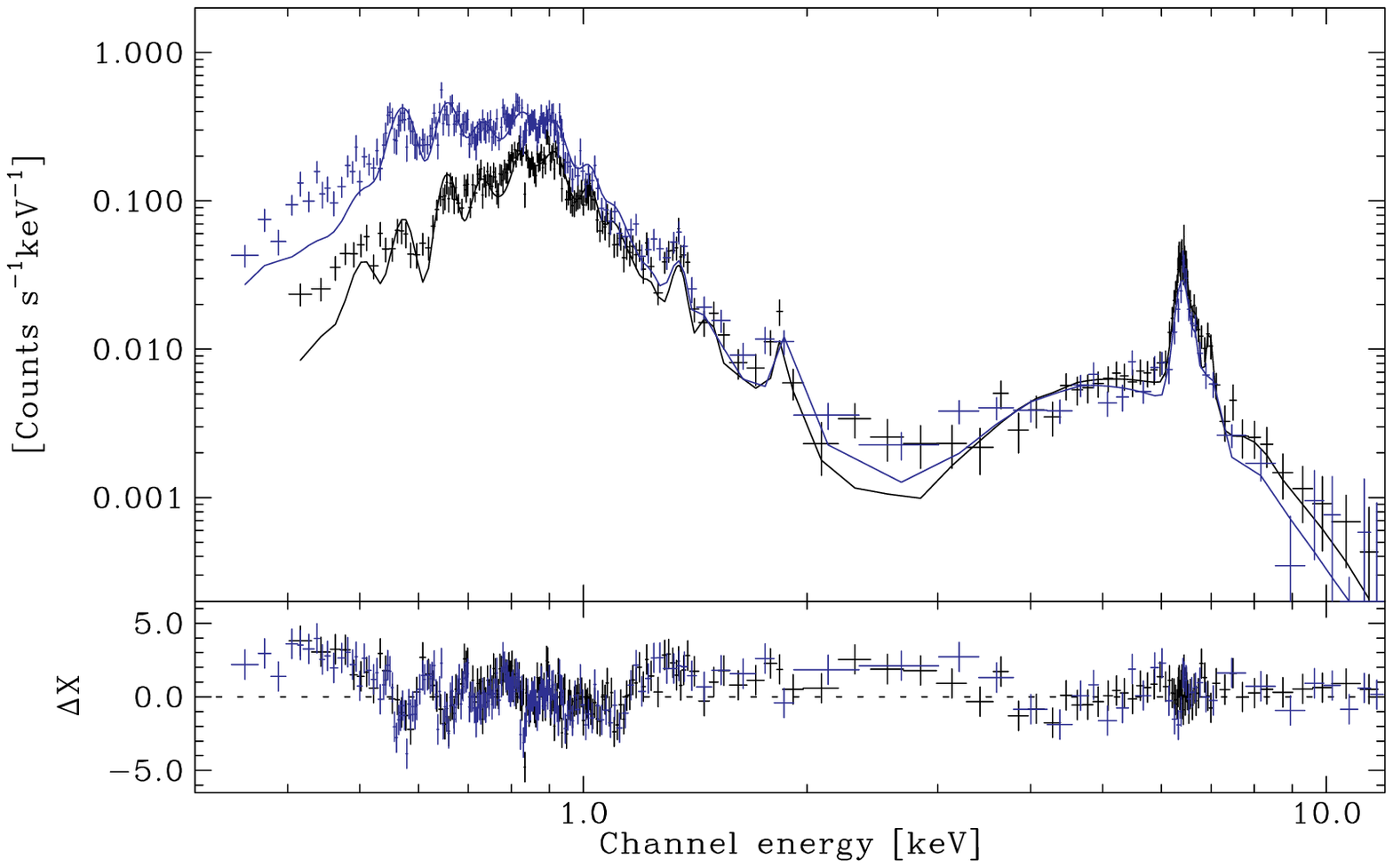}
  \end{center}
  \caption{{\sl Suzaku\/} XIS spectra of CH Cyg in 2006 January (top)
           and in 2006 May (bottom). The observed data are shown
	   together with a model (see text for a full description)
	   folded through the instrument response.}\label{fig:xis}
\end{figure*}

We present the average XIS spectra of CH Cyg from the two
{\sl Suzaku\/} observations in Figure\,\ref{fig:xis}.  Data
for the FI chips (black) and the BI chip (blue) are shown with
a model consisting of a hard (kT$\sim$10 keV) thermal component
(the mekal model in XSPEC) seen through a simple absorber (3.2 and 2.2 $\times
10^{23}$ cm$^{-2}$ respectively, for the two observations), plus a soft
component represented as a two-temperature (0.2, 0.6 keV) mekal model.
We have included an interstellar absorption term
that applies to both the soft and the hard components, but it
is not significantly detected with {\sl Suzaku\/}.

We do not consider this to be the best model; for example,
two-temperature plasma is likely just a convenient representation
of the soft component, while the extra
absorber that the hard component goes through probably requires
a complex (such as partial covering) absorber.  However, the
accurate characterization of the latter is hampered by the relatively
narrow effective bandpass (this component is detected strongly between
3 and 10 keV, and one has to worry about the contribution of the
soft component near 3 keV).  As for the soft component, we have
not attempted a more sophisticated fit partly because of the remaining
calibration uncertainties in the amount of contamination, as well
as small residual problems in gain and resolution.  Of these,
allowing for an extra line broadening in particular ($\sigma$=14 eV
for the 2006 Jan observation) significantly reduces the structured
residuals between 0.5 and 1 keV.  In addition to calibration issues,
the use of such a relatively simple model better facilitates a direct
comparison with the {\sl ASCA\/} and {\sl Chandra\/} data.

\begin{figure*}
  \begin{center}
    \FigureFile(139.5mm,116.4mm){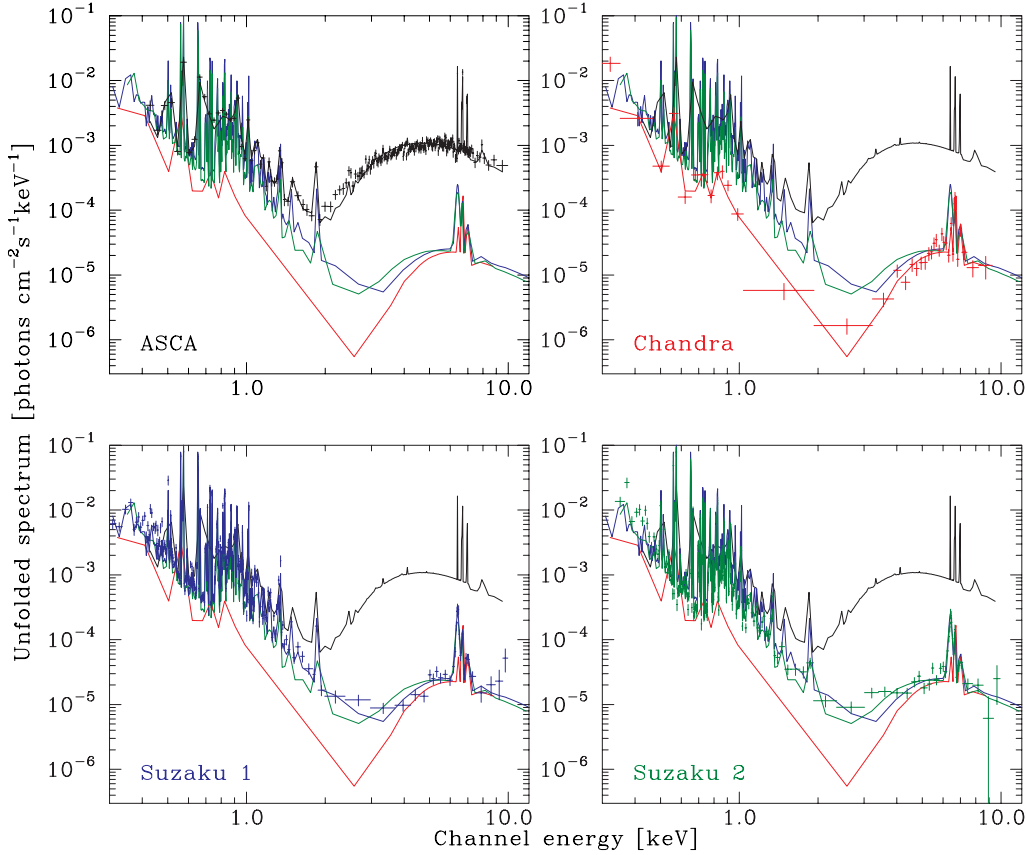}
  \end{center}
  \caption{Unfolded spectra from the {\sl ASCA\/}, {\sl Chandra\/}
           (red) and {\sl Suzaku\/} (blue and green) observations.
           Each panel shows the unfolded spectrum (data and model)
	   for one observation, as well as the best-fit model for
	   the other three observations for ease of comparison.
	   See text for details of the spectral model.}\label{fig:unf}
\end{figure*}

In the folded representation, these {\sl Suzaku\/} spectra may
look similar to the {\sl ASCA\/} spectrum already published
by \citet{EIM1997}.  However, it turns out that, while the
soft X-ray fluxes are similar, the hard X-ray flux levels
observed in the 2006 observations are much lower than that seen in 1994.
In contrast, during the {\sl Chandra\/} observation, both the
soft and the hard components were much weaker than during the
{\sl ASCA\/} observations.  We present the 0.4--2 keV and 3--10 keV
fluxes in these 4 observations (without correcting for the strong
absorption of the hard component) in Table\,\ref{tab:obslog}
and the unfolded spectra in Figure\,\ref{fig:unf} to demonstrate
the dramatic changes.

\begin{figure*}
  \begin{center}
    \FigureFile(148.8mm,121.92mm){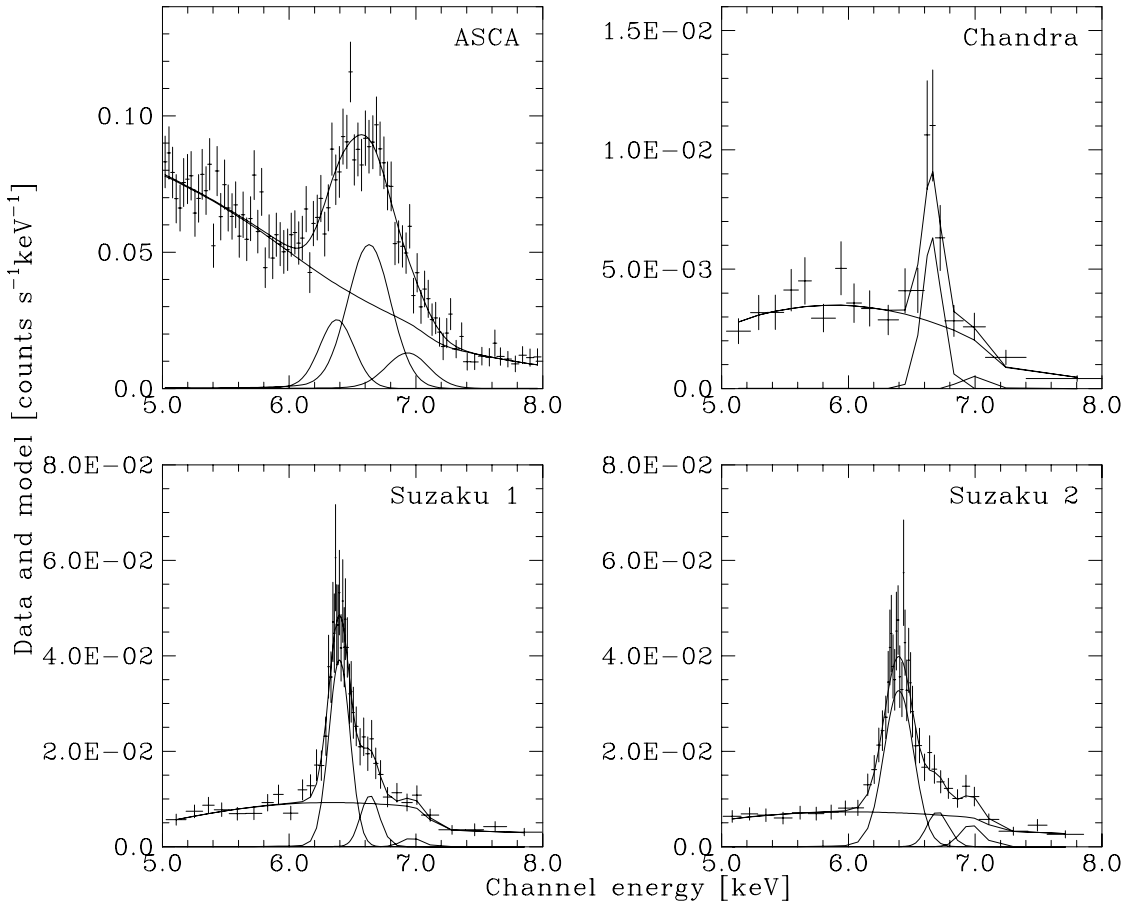}
  \end{center}
  \caption{The Fe K$\alpha$ profiles of CH Cyg from the 4 observations.
           {\sl ASCA\/} SIS, {\sl Chandra\/} ACIS-S (zeroth order),
	   and {\sl Suzaku} FI (3 units combined) data are shown with
	   absorbed bremsstrahlung plus Gaussian fits.}\label{fig:fek}
\end{figure*}

A detailed study of the Fe K$\alpha$ lines (Figure\,\ref{fig:fek}) reveals
another dramatic change.  For this, we have fitted the spectra in the 5--10
keV region with an absorbed bremsstrahlung continuum plus up to three
Gaussians.  The center energy of the fluorescent and hydrogenic components
were fixed at 6.4 and 6.97 keV, respectively, while the energy of the helium
like line was allowed to float around  6.7 keV.  The widths were initially
set to 0, but were then allowed to vary, keeping those of the  He-like and
H-like lines to be identical\footnote{Due to the current limitations in
the calibration of XIS response, we refrain from any interpretations
of the apparent line width in our {\sl Suzaku\/} data.  In particular,
the apparent broadening from the first observation to the second may be
due to changes in the instrumental response that is yet to be modeled.}.
Some of these restrictions are necessary for
the fit to converge, but we caution that the fit results are therefore not
assumption-free.  Nevertheless, it is clear that the Fe K$\alpha$ complex
during the {\sl Suzaku\/} observations are dominated by the fluorescent
line, with equivalent widths of 0.7--1.3 keV (this range representing
the results of fitting the complex with slightly different models, although
we cannot guarantee that we have tried all possible models).  This
component was not detected in the {\sl Chandra\/} data, while the
{\sl ASCA\/} spectrum appears to contain a more normal (Eq.W$\sim$140 eV)
fluorescent line.  We note that the decomposition of the complex in
the {\sl ASCA\/} data is not unique, and a solution without a fluorescent
line is also possible.  Derived parameters are summarized in
Table\,\ref{tab:fek}.

\begin{figure*}
  \begin{center}
    \FigureFile(83.7mm,70.38mm){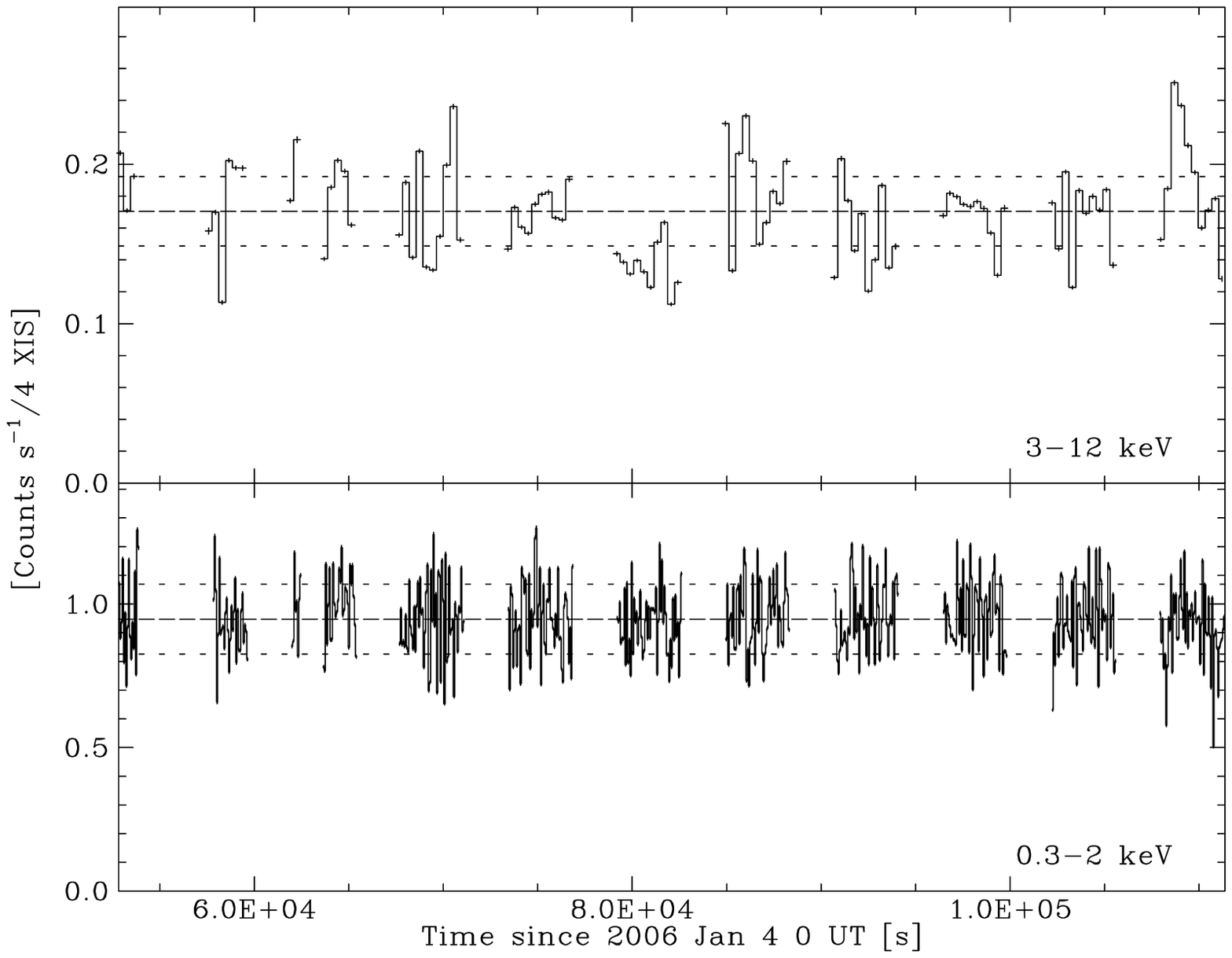}
    \FigureFile(83.7mm,70.38mm){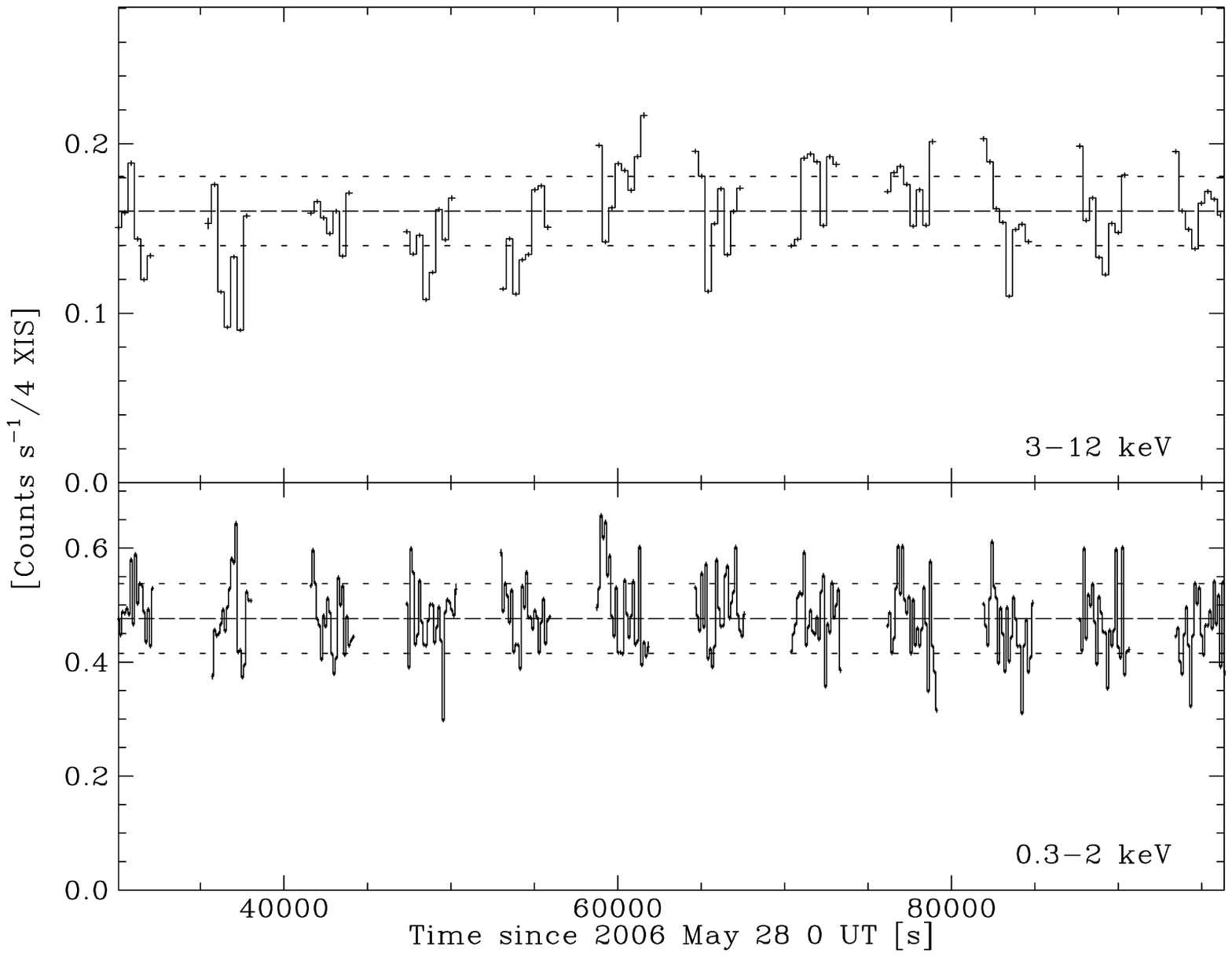}
  \end{center}
  \caption{{\sl Suzaku\/} XIS light curves (summed count rates in all 4 XIS
           units) of CH Cyg in 2006 January (left) and in 2006 May (right),
           plotted as histograms.
	   The bin sizes are 64 s (soft band) and 360 s (hard)
	   for the January data, and 128 s (soft) and 384 s (hard) for the May
	   data, resulting in 60 source counts in each bin on average.
	   The global average and $\pm 1 \sigma$ levels are indicated by
	   dashed lines. The difference in the mean count rate in the soft
           band are due both to the contamination build-up and to a change
           in the source flux.}\label{fig:lc}
\end{figure*}

We have also analyzed the variability of the soft and hard components.
We note, however, that there is a  30--40$''$ level attitude wobble
that is not taken into account in the current attitude solution, which
may well introduce spurious variations in the light curves.  We have
adjusted the bin sizes for the two bands and in two observations such
that, on average, each bin has 60 source counts, or a single bin
signal-to-noise ratio of $\sim$7.8, and display the results in
Figure\,\ref{fig:lc}.  The hard-band light curves,
particularly those of the January data, appear to be significantly
more variable than the soft-band curves, showing clusters of points
with correlated deviation from the mean.  For example, the dip around
80,000 s and the flare-like event around 120,000 s in the January data
are significantly more than the statistical fluctuation in
the hard band, but are not seen in the the soft band.  Overall, in 2006
January, the mean count rate in the soft band is 0.948 cts\,s$^{-1}$;
for 64 s bins, the expected statistical error is 0.122 cts\,s$^{-1}$
(1$\sigma$), while the actual observed standard deviation is 0.126
cts\,s$^{-1}$.  For the mean in the hard band is 0.170 cts\,s$^{-1}$,
the statistical error is 0.022 cts\,s$^{-1}$ (1$\sigma$ in 360 s bins),
the observed standard deviation is 0.030 cts\,s$^{-1}$.  That is,
the hard band light curve shows a larger scatter than expected
on a purely statistical basis, while the scatter in the soft band
light curve is only marginally above the expected level at most.  Similarly,
in 2006 May, the numbers (mean, observed standard deviation, and expected
1$\sigma$ statistical error) are 0.477, 0.0622, and 0.610 cts\,s$^{-1}$
in the soft band (128 s bin), and 0.158, 0.026, and 0.020 cts\,s$^{-1}$
in the hard band (384 s bin).  We therefore confirm that the hard component
is more variable than the soft component \citep{EIM1997}.

The mean count rate in the soft band changed from 0.95 cts\,s$^{-1}$
in 2006 January to 0.48 cts\,s$^{-1}$ in 2006 May.  Although the
contamination build-up plays a significant role in this, the soft
flux of CH~Cyg does appear to have declined by roughly 30\% over
this period (Table\,\ref{tab:obslog}).  There is little doubt as to
the reality of the longer-term variability of the soft component
(\cite{LT1987}; Table\,\ref{tab:obslog}; Figure\,\ref{fig:unf}).

\section{Discussion}

\begin{table*}
  \caption{Fe K$\alpha$ Lines of CH Cyg.}\label{tab:fek}
  \begin{center}
    \begin{tabular}{lllll}
      Observation & 6.4 keV Eq.W. & 6.7 keV Eq.W & 6.7 keV Energy & 6.97 keV Eq. W \\
       & (eV) & (eV) & (keV) & (eV) \\
      {\sl ASCA} & 140 & 500 & 6.66 (6.57--6.71)  & 140 \\
      {\sl Chandra} & - & 390 & 6.66 (6.63--6.69)  & 30 \\
      {\sl Suzaku\/}-1 & 690 & 90 & 6.64 (6.60--6.68) & 40 \\
      {\sl Suzaku\/}-2 & 1150 & 90 & 6.71 (6.66--6.81) & 120 \\
    \end{tabular}
  \end{center}
\end{table*}

CH Cyg has shown a dramatic spectral variability over the last dozen years
or so, while keeping the same basic two-component structures.  At the
same time, {\sl Swift\/} BAT 14--25 keV light curve of CH Cyg also shows
a decline, from $\sim 1.0 \times 10^{-4}$ cts\,s$^{-1}$ in the 14--25 keV
band during the first 3 months of 2005 to an undetectable level
($< 2.0 \times 10^{-5}$ cts\,s$^{-1}$) by the end of 2005  \citep{Kea2006}.
We first consider the nature of the hard X-ray decline from the
{\sl ASCA\/} era to the {\sl Suzaku\/} era.

The extraordinary strength of the 6.4 keV, fluorescent Fe K$\alpha$ line
shows that our direct line-of-sight is now blocked and we only observe
hard X-ray photons scattered into our line of sight \citep{I1985}.
The low-energy cut-offs of the hard component did not vary much
among the four observations shown in Figure\,\ref{fig:unf}.
The best-fit N$_{\rm H}$ values were 1.4, 4.1, 3.2, and 2.2
$\times 10^{23}$ cm$^{-2}$, respectively in our fits to the
four observations.  Such levels of absorption, which are admittedly
uncertain due to the relatively simple model we have adopted,
are insufficient to explain the observed equivalent widths of the
6.4 keV line in Case II of \authorcite{I1985} (spherically symmetric
absorber, direct and scattered X-rays both observed at Earth),
and requires Case III (direct X-rays hidden).

In contrast, the Fe K$\alpha$ complex was dominated by the thermal
lines in the {\sl ASCA\/} data, so in that observation our direct
line of sight was open.  The continuum flux declined by a factor
of $\sim$30.  In contrast, the 6.4 keV line fluxes declined from
3.5$\times 10^{-4}$ photons\,cm$^{-2}$s$^{-1}$ to 1.4 and 1.0$\times 10^{-4}$
photons\,cm$^{-2}$s$^{-1}$, respectively, in the two {\sl Suzaku\/}
observations.  It is therefore likely that all of the direct
X-rays, as well as the majority (60--70\%) of the scattered
X-rays, are blocked during the {\sl Suzaku\/} era.

% The simple extrapolation of kT$\sim$10 keV thermal spectrum up to
% 25 keV suggests that the {\sl ASCA\/} hard X-ray flux level is
% compatible with the {\sl Swift\/} BAT signal during the first half
% of 2005.  It also shows that the {\sl Suzaku\/} HXD/PIN detection
% is at a higher flux than expected from a simple extrapolation of the
% XIS data.  However, scattered component should have a prominent
% Compton hump, similar to that shown by reflection component,
% so such an excess is to be expected.

As for the central engine that produces the hard X-ray emission,
we consider it likely that this is due to accretion onto the white
dwarf, following \citet{EIM1997}.  In fact, the temperature
(kT$\sim$10 keV) and the luminosity (10$^{33}$ ergs\,s$^{-1}$)
seen by {\sl ASCA\/} are both similar to that seen in the old nova,
V603~Aql \citep{MO2005}. If this analogy is correct, then the X-rays
are probably generated in the boundary layer between the accretion
disk and a relatively massive white dwarf.

Between the M giant wind and the corona, wind, and/or jet from the
accretion disk, there are more than enough candidates for the scattering
medium.  However, as \citet{vdBea2006} have calculated, typical
line-of-sight column through the M giant wind in a symbiotic system
is $<2 \times 10^{21}$ cm$^{-2}$, which is borne out observationally
by the widespread detection of the supersoft X-rays in symbiotic stars
\citep{MWJ1997}.  M giant winds are therefore far too optically thin
to produce a significant amount scattered X-rays.  Unless the white
dwarf in CH Cyg is unusually close to the photosphere of the
M giant, which now seems unlikely with the preference for a 14.5 yr
orbital period \citep{Sea2006}, or the M giant has much stronger
mass loss than in other symbiotic systems, the M giant wind
is therefore unlikely to be the site of hard X-ray absorption
and scattering.  Moreover, the strong decrease in the 6.4 keV
flux also points to a relatively compact site for the scattering,
i.e., material associated with the accretion disk, such as
an accretion disk corona or an accretion disk wind.

What about the Compton-thick material that is required to
block out the direct hard X-rays?  The M giant itself is a
candidate.  However, according to \citet{Sea1996}, the eclipse
of the 14.5 yr binary would have happened around 2000\footnote{In
addition, the ephemeris for the 756 day ``binary'' indicates that
the white dwarf would have been near inferior conjunction in 2006
January through May.}.  Thus, the existing orbital solution of CH Cyg
suggests that the 2006 low state is not due to an eclipse by the
M giant.

We therefore consider it likely that the blockage was caused
by the accretion disk.  Recall that jet observations indicate
a high inclination angle for the disk \citep{S1987}, very close
to 90$^\circ$.  Perhaps the disk thickness varies from epoch
to epoch; perhaps the disk is slightly tilted and precesses.
In either case, it seems reasonable for a highly inclined disk
to be able to block our direct line-of-sight.  If the scattering
medium is, say, a relatively compact accretion disk corona, it is
plausible for a tilted and/or thickened outer disk also to block
a significant fraction of the scattered X-rays, as we have inferred
from the decline in fluorescent Fe line flux.

If this picture is correct, the same thickening or precession
of the accretion disk may also explain some of the optical/UV
variability of CH Cyg.  New optical/UV observations in the
{\sl Swift\/} era will enable a direct comparison of hard X-ray
and optical/UV variabilites.

What can we learn about the soft component from the {\sl Suzaku\/}
data?  We believe we now have sufficient data to disprove the
ionized absorber interpretation of \citet{W2001}.  First, the
analysis of \citet{EIM1997} showed that while the hard component
was significantly variable, the soft component was consistent
with Poisson noise.  This is difficult to explain in the ionized
absorber model.  Second, we do not believe that the Compton-thick
material blocking our line of sight to the hard X-ray emission region
can be significantly ionized, given the modest X-ray
luminosity (of order 10$^{33}$ ergs\,s$^{-1}$) of CH Cyg.
Finally, the He-like line of Mg is clearly seen in Figure\,\ref{fig:xis},
while the H-like line is not.  Since the latter is stronger in
any plasma hotter than kT$\sim$0.8 keV, we conclude that the soft
component is dominated by plasma cooler than kT$\sim$0.8 keV.
This is also consistent with the strengths of other emission lines
(O, Fe L, Ne) seen below 1 keV, which would be significantly diluted
if much hotter plasma co-existed in the emission region.

We therefore conclude that the soft X-rays indeed are a distinct
component.  The emitting region must be separate from, or significantly
more extended than, the hard X-ray emitting region.  On the other hand,
the bulk of the soft X-rays were spatially unresolved with {\sl Chandra\/}
\citep{GS2004}.  This sets the maximum size to be of order 100 AU
(0.4$''$ at 250 pc), or 800 light minute in radius.  The entire
binary is likely to be smaller than this, assuming a 14.5 yr orbital
period.  An extended emission region, comparable to the size of
the binary, is expected in the colliding wind interpretation,
in which the M giant wind is colliding with an outflow from
the vicinity of the accreting white dwarf. The exact nature of the latter
is unclear; it can be a compact jet (unlike that produced
the spatially-resolved X-rays observed with {\sl Chandra\/}; \cite{GS2004}),
or it can be a less collimated outflow, more appropriately termed wind.
Colliding wind systems are usually not highly variable on short
time scales, presumably because any local instabilities are averaged
over the large emission site.  Only a global change (changes in binary
separation or in the wind mass loss rate) can modulate the X-ray
flux significantly.  If the colliding wind emission is from a 10 AU
size region, and the accretion disk wind velocity is 1,000 km\,s$^{-1}$,
then we do not expect significant modulation in soft X-rays on
time scales shorter than about 2 weeks from a ``wind travel time''
argument.  This consistent with the lack of soft X-ray variability
seen with {\sl ASCA\/} \citep{EIM1997}.

\citet{EIM1997} considered the soft component to be from the
corona of the M giant.  However, M giants are in the ``coronal
graveyard'' to the upper right of the dividing line in the HR diagram.
{\sl Chandra\/} observations of single giants in this region
show that coronal emissions are far too low a level to explain
the soft component of CH Cyg: Arcturus is detected at 3$\sigma$ level
at a 0.2--2 keV luminosity of  $\sim 1.5 \times 10^{25}$ ergs\,s$^{-1}$
and the upper limit for Aldebaran is 7$\times 10^{25}$ \citep{ABH2003};
the M supergiant Betelgeuse has a {\sl Chandra\/} upper limit of
$\sim 10^{28}$ ergs\,s$^{-1}$ for coronal temperature $>10^6$ K
\citep{PBea2006}.  Whenever M giants are detected in X-rays,
they are usually interpreted as due to a binary companion
(\cite{vdBea2006} and references therein).  Although binarity
can affect the X-ray luminosity, this is limited to short period
systems (periods of 2 days -- 2 weeks for RS CVn systems, for example)
where tidal torque would lead to synchronous rotation, which is
unlikely in a symbiotic binary.  For the coronal model of the
soft X-ray emission to succeed, the mass donor in the CH Cyg system
must be a truly exceptional M giant.

So far, we have concentrated on the comparison between {\sl ASCA\/}
and {\sl Suzaku\/} data.  What about the {\sl Chandra\/} observation,
during which both the soft and the hard X-ray components were fainter
and there was no detection of the fluorescent Fe K$\alpha$ line?
For this, we must invoke a separate mechanism.  If the accretion rate
onto the white dwarf declines dramatically, we expect both the hard
and the soft X-ray luminosities to drop, and this appears to be what
happened in 2001.

\section{Conclusions}

Our two {\sl Suzaku\/} observations of CH Cyg have caught the system
in a soft X-ray bright, hard X-ray dim state.  Based on the Fe K$\alpha$
spectra, we believe we currently observe only the scattered hard X-ray
component.  The decline in the hard X-ray flux is only an apparent
decline, and the intrinsic X-ray luminosity of CH Cyg may still be
as high as during the {\sl ASCA\/} observation, though the actual value
is unknowable.

A symbiotic star such as CH Cyg is filled with possible scattering media,
but we favor material above the accretion disk as the scattering site.
As for the object that blocks our direct view of
the hard X-ray emission, we believe it is more likely to be an edge-on
accretion disk than the M giant mass donor.  The hard X-rays presumably
originate from accretion onto the white dwarf, but further progress would
require high-quality X-ray spectroscopy when the direct view is not blocked.

\subsection*{Acknowledgements}

We thank Drs. M. Corcoran and S. Drake for useful discussion on
colliding winds and M giant coronae.


\newpage

\begin{thebibliography}{}

\bibitem[Ayres et al. (2003)]{ABH2003}
  Ayres, T.R., Brown, A., Harper, G.M. \ 2003, \apj, 598, 610
% Coronal Graveyard

\bibitem[Ezuka et al. (1997)]{EIM1997}
  Ezuka, H., Ishida, M., Makino, F. \ 1997, \apj, 499, 388
% ASCA

\bibitem[Galloway \& Sokoloski (2004)]{GS2004}
  Galloway, D.K., Sokoloski, J.L. \ 2004, \apj, 613, L61
% X-ray jet

\bibitem[Hinkle et al. (1993)]{Hea1993}
  Hinkle, K.H., Fekel, F.C., Johnson, D.S., Scharlach, W.W.G. \ 1993,
  \aj, 105, 1074
% Triple system claimed

\bibitem[Inoue (1985)]{I1985}
  Inoue, H. \ 1985, Sp.Sci.Rev, 40, 317
% Fe Kalpha

\bibitem[Kennea et al. (2006)]{Kea2006}
  Kennea, J.A., Mukai, K., .... \ 2006, in preparation
% BAT Symbiotic paper

\bibitem[Kokubun et al. (2006)]{HXDPER2006}
  Kokubun, M. et al. \ 2006, \pasj, in press
% HXD Performance

\bibitem[Koyama et al. (2006)]{XIS2006}
  Koyama, K. et al. \ 2006, \pasj, in press
% XIS

\bibitem[Leahy \& Taylor (1987)]{LT1987}
  Leahy, D.a., Taylor, A.R. \ 1987, \aap, 176, 262
% Early X-ray emission

\bibitem[Mitsuda et al. (2006)]{SUZ2006}
  Mitsuda, K. et al. \ 2006, \pasj, in press
% Suzaku mission reference

\bibitem[Mukai \& Orio (2005)]{MO2005}
  Mukai, K., Orio, M. \ 2005, \apj, 622, 602
% V603 Aql

\bibitem[Munari et al. (1996)]{Mea1996}
  Munari, U., Yudin, B.F., Kolotilov, E.A., Tomov, T.V. \ 1996, \aap, 311, 484
% Triple model doubted

\bibitem[M\"urset et al. (1997)]{MWJ1997}
  M\"urset, U., Wolff, B., Jordan, S. \ 1997, \aap, 319, 201
% ROSAT survey of symbiotic stars

\bibitem[Perryman et al. (1997)]{Pea1997}
  Perryman, M.A.C., et al. \ 1997, \aap, 176, 262
% Hipparcos Parallax

\bibitem[Posson-Brown et al. (2006)]{PBea2006}
  Posson-Brown, J., Kashyap, V.L., Pease, D.O., Drake, J.J. \ 2006,
  \apj, submitted (astro-ph/0606387)
% Betelgeuse upper limit

\bibitem[Schmidt et al. (2006)]{Sea2006}
  Schmidt, M.R., Zacs, L., Mikolajewska, J., Hinkle, K.H. \ 2006,
  \aap, 446, 603
% Hinkle now agrees 756 day period is due to pulsation

\bibitem[Serlemitsos et al. (2006)]{XRT2006}
  Serlemitsos, P.J. et al. \ 2006, \pasj, submitted.
% XRT

\bibitem[Skopal et al. (1996)]{Sea1996}
  Skopal, A., Bode, M.F., Lloyd, H.M., Tamura, S. \ 1996, \aap, 308, L9
% Eclipses ?!?

\bibitem[Solf (1987)]{S1987}
  Solf, J. \ 1987, \aap, 180, 207
% O[III] jet

\bibitem[Takahashi et al. (2006)]{HXD2006}
  Takahashi, T. et al. \ 2006, \pasj, submitted.
% HXD

\bibitem[Taylor et al. (1986)]{TSM1986}
  Taylor, A.R., Seaquist, E.R., Mattei, J.A. \ 1986, \nat, 319, 38
% Radio jet

\bibitem[van den Berg et al. (2006)]{vdBea2006}
  van den Berg, M., Grindlay, J., Laycock, S., Hong, J., Zhao, P., Koenig, X.,
  Schlegel, E.M., Cohn, H., Lugger, P., Rich, R.M., Dupree, A.K., Smith, G.H.,
  Strader, J. \ 2006, \apjl, 647, L135
% ChamPLANE Symbiotics

\bibitem[Wheatley (2001)]{W2001}
  Wheatley, P.J. \ 2001, AIP Conf. Proc. 599, X-ray Astronomy:
  Stellar Endpoints, AGN, and the Diffuse X-ray Background (Melville: AIP),
  1007
% Ionized absorber interpretation.

\end{thebibliography}
\end{document}